\documentstyle[11pt,newpasp,epsf]{article}
\begin{document}

\title{Investigating a Possible Spectral Signature of the Wind-ISM
  Interaction Region of $\alpha$~Tau}

\author{Brian E. Wood}
\affil{JILA, University of Colorado, Boulder, CO 80309-0440}
\author{Graham M. Harper}
\affil{Center for Astrophysics and Space Astronomy, University of Colorado,
  Boulder, CO 80309-0389}
\author{Hans-Reinhard M\"{u}ller\altaffilmark{1}}
\affil{Bartol Research Institute, University of Delaware, Newark,
  DE 19716}
\author{Gary P. Zank}
\affil{Institute of Geophysics and Planetary Physics,
   University of California at Riverside, 1432 Geology, Riverside, CA 92521}

\altaffiltext{1}{Also at IGPP, University of California at Riverside,
  1432 Geology, Riverside, CA 92521}

\begin{abstract}

     Ultraviolet spectra from the GHRS instrument on board the
{\em Hubble Space Telescope} reveal the presence of a mysterious
absorption feature in the Mg~II h \& k lines of the nearby ($d=20.0$~pc)
K5~III star $\alpha$~Tau.  The narrow absorption looks like an
interstellar absorption feature but it is in the wrong location based on
our knowledge of the local ISM flow vector.  Since the absorption is
close to the rest frame of the star, it has been interpreted as being from
the interaction region between $\alpha$~Tau's massive, cool wind and the
interstellar medium, i.e., $\alpha$~Tau's ``astrosphere''.  We compute
hydrodynamic models of the $\alpha$~Tau astrosphere in order to see if the
models can reproduce the Mg~II absorption feature.  These models do
predict that stellar wind material heated, decelerated, and compressed
after passing through a termination shock a few thousand AU from the star
should produce a Mg~II absorption feature with about the right width at
roughly the right velocity.  However, our first models underestimate
the Mg~II column density by an order of magnitude.  A much larger
parameter search is necessary to see whether the observed Mg~II absorption
can be reproduced by acceptable changes to the adopted stellar
wind and ISM properties.

\end{abstract}

% Keywords should be included, but they are not printed in the hardcopy.
% They will be used by the Editors to help organize poster papers by
% category though!


\index{*Alpha Tau}
\index{Hubble Space Telescope}
\index{Ultraviolet}

\section{The Mysterious Mg~II Absorption Feature}

     Figure~1 shows the Mg~II k line from $\alpha$~Tau
(K5~III) observed by the Goddard High Resolution Spectrograph (GHRS) on
board the {\em Hubble Space Telescope} (HST) on 1994 April 8, first studied
by Robinson et al.\ (1998).  The center of the stellar line profile is
dominated by very broad absorption from the massive, cool wind
of the red giant star.  However, near the rest frame of the star there is
a much narrower absorption feature, which is also seen in the Mg~II h line.
The narrow absorption looks very much like interstellar absorption, and
the strength of the absorption in the k and h lines has the expected 2 to
1 ratio of the oscillator absorption strengths, but Robinson et al.\ (1998)
noted that the location of the absorption makes an interstellar origin
very unlikely.
\begin{figure}[t]
\plotfiddle{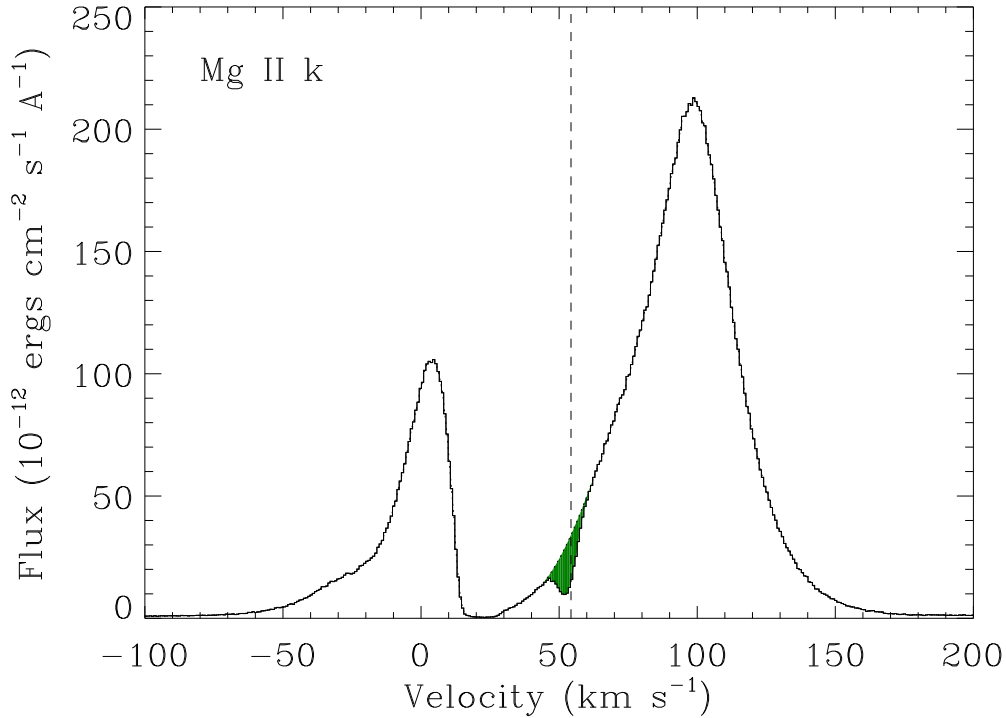}{3.6in}{90}{60}{60}{233}{-10}
\caption{HST/GHRS Mg~II k line spectrum of $\alpha$~Tau, plotted
  on a heliocentric velocity scale, with the rest frame of the star
  indicated by the vertical dashed line.  Broad absorption from
  $\alpha$~Tau's wind dominates the central part of the profile.  The
  narrow absorption feature shaded in green is the absorption
  that is presumed to be from the wind/ISM interaction region.}
\end{figure}

     The flow vector of the Local Interstellar Cloud (LIC) from Lallement
et al.\ (1995) suggests that LIC absorption should be at
$+25.5$ km~s$^{-1}$, which is within the saturated core of the wind
absorption in Figure~1, and is therefore undetectable. 
Interstellar absorption components other than that of the LIC are
occasionally seen for stars as nearby as $\alpha$~Tau ($d=20.0$~pc), but
they are generally close to the expected LIC velocity.  Redfield \&
Linsky's (2002) survey of ISM absorption features within 100~pc
demonstrates just how discrepant the location of the $\alpha$~Tau
absorption feature is, if it is interstellar.

     Noting the absorption feature's location near the stellar rest frame,
Robinson et al.\ (1998) proposed that the absorption is instead from the
interaction region between $\alpha$~Tau's wind and the ISM.  The absorption
would therefore be somewhat analogous to the H~I Ly$\alpha$ absorption
detected from the interaction regions between the winds of solar-like stars
and the local ISM (Linsky \& Wood 1996; Wood et al.\ 2000, 2001, 2002;
M\"{u}ller et al.\ 2001).  This ``astrospheric'' (analogous with
``heliospheric'') absorption has proved to be a very useful diagnostic for
the wind properties of solar-like stars (Wood et al.\ 2002).

\section{Modeling the $\alpha$~Tau Astrosphere}

     We modeled the $\alpha$~Tau wind/ISM interaction region in order to
see whether $\alpha$~Tau's astrosphere can indeed account for the narrow
Mg~II absorption feature.  Table~1 lists the ISM parameters and stellar
wind properties at 1~AU from the star that are assumed
in the initial astrospheric model (Model~1).  The stellar wind parameters
are estimated from an analysis of the Mg~II wind absorption profile (e.g.,
Harper et al.\ 1995; Robinson et al.\ 1998), which suggests a mass loss rate
of $1.0\times 10^{-11}$ M$_{\odot}$ yr$^{-1}$, a result also consistent with
an upper limit derived from VLA radio data.  However, significant
uncertainties exist for all the wind parameters in Table~1 and in Model~2
we compute a model with a significantly higher mass loss rate by
increasing the assumed wind densities, $n_{w}({\rm H~I})$ and
$n_{w}({\rm H^{+}})$, by a factor of 10.

\begin{table}[t]
\caption{Model Input Parameters}
\begin{center}
\scriptsize
\begin{tabular}{cccccccccc}
\tableline
\# & \multicolumn{4}{c}{Stellar Wind Parameters at 1 AU} &
  \multicolumn{5}{c}{ISM Parameters} \\
 & $n_{w}({\rm H~I})$ & $n_{w}({\rm H^{+}})$ & $V_{w}$ & $T_{w}$ &
  $n_{\infty}({\rm H~I})$ & $n_{\infty}({\rm H^{+}})$ & $V_{\infty}$ &
  $T_{\infty}$ & $\theta$ \\
 & (cm$^{-3}$) & (cm$^{-3}$) & (km s$^{-1}$) & (K) &
  (cm$^{-3}$) & (cm$^{-3}$) & (km s$^{-1}$) & (K) & (deg) \\
\tableline
1 & $3\times 10^{4}$ & $5\times 10^{3}$ & 27 & 7500 &
  0 & $4\times 10^{-3}$ & 34 & $5\times 10^{5}$ & 149 \\
2 & $3\times 10^{5}$ & $5\times 10^{4}$ & 27 & 7500 &
  0 & $4\times 10^{-3}$ & 34 & $5\times 10^{5}$ & 149 \\
3 & $3\times 10^{4}$ & $5\times 10^{3}$ & 27 & 7500 &
  0 & $4\times 10^{-3}$ & 44 & $5\times 10^{5}$ & 170 \\
\tableline
\end{tabular}
\end{center}
\end{table}

     Unlike the Sun and other coronal stars with detected
astrospheric absorption (e.g., Wood et al.\ 2002), $\alpha$~Tau probably
lies in the hot ISM rather than the warm, neutral ISM.  This conclusion is
based on observations that show that in the direction of $\alpha$~Tau
($l=181^{\circ}$, $b=-20^{\circ}$) H~I column densities do not increase much
with distance once the line of sight is $>5$~pc from the Sun, meaning that
beyond $\sim 5$~pc the ISM is hot and ionized and therefore contains no H~I
(Piskunov et al.\ 1997; Redfield \& Linsky 2000, 2001).  Such hot material
fills most of the Local Bubble and is believed to account for much of the
soft X-ray background (Sfeir et al.\ 1999).

     The exact temperature of the hot ISM within the Local
Bubble is not precisely known, but it is believed to be $\sim 10^{6}$~K.
In our models, we assume $T_{\infty}=5\times 10^{5}$~K (see Table~1),
and we assume a proton density such that the pressure is about the same as
that known to exist for the ISM around the Sun.  Although the flow vector
for the LIC is well known, and it is also known that other warm neutral
clouds near the Sun must have similar vectors, it is uncertain whether the
LIC vector will apply to the hot ISM material within the Local Bubble.
Nevertheless, for Models 1 and 2 we have assumed the LIC vector in deriving
the ISM flow velocity seen by the star ($V_{\infty}=34$ km s$^{-1}$).

     For Model~3, we assume that the hot ISM is at rest in the Local
Standard of Rest (LSR).  The Local Bubble was presumably
created by a series of supernova explosions (Ma\'{i}z-Apell\'{a}niz 2001;
Bergh\"{o}fer \& Breitschwerdt 2002).  Since the hot stars that produce
supernovae generally have low proper motions relative to the LSR, the
assumption that the hot ISM within the Local Bubble is at rest in the LSR
is a plausible one.  Based on this assumption, the ISM flow velocity seen
by the star increases to $V_{\infty}=44$ km s$^{-1}$.
Assuming the LIC vector, the line of sight from the star toward the Sun
is $\theta=149^{\circ}$ from the upwind direction of the ISM flow
seen by $\alpha$~Tau (see last column of Table~1), which means we are
looking at the downwind part of the $\alpha$~Tau astrosphere.  This is
even more true with the LSR assumption, where $\theta=170^{\circ}$, only
$10^{\circ}$ from directly downwind.

     In modeling $\alpha$~Tau's astrosphere, we use the same hydrodynamic
codes used to model the heliosphere and solar-like astrospheres
(Pauls et al.\ 1995; Zank et al.\ 1996; Wood et al.\ 2000, 2002;
M\"{u}ller et al.\ 2001).  We initially tried to use the ``four-fluid''
code of Zank et al.\ (1996), where the wind/ISM interaction is
represented as the interaction between one plasma fluid and three separate
neutral H fluids.  However, we had trouble getting this code to converge,
so we switched to a simpler ``two-fluid'' code like that of Pauls et al.\
(1995), where the neutrals are only described by a single fluid component.
Based on this code, Figure~2 shows density
distributions, temperature distributions, and flow patterns for the protons
and H atoms for Model~1.
\begin{figure}[p]
\plotfiddle{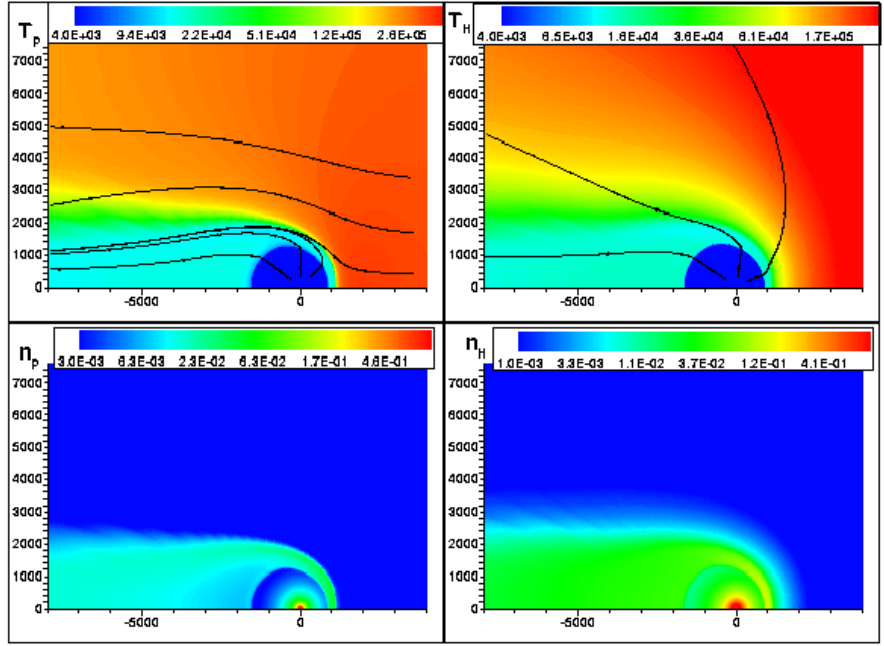}{3.1in}{0}{57}{57}{-175}{-113}
\caption{Hydrodynamic model of the $\alpha$~Tau astrosphere for Model~1,
  where the upper panels are the proton and H~I temperature and the bottom
  panels are the proton and H~I density, respectively.  Streamlines are
  shown in the upper panels.  The distance scale is in AU.}
\end{figure}

     The astrospheric structure of $\alpha$~Tau is in many respects
similar to that of the heliosphere (see, e.g., Zank et al.\ 1996). The
stellar wind expands radially until it reaches a termination shock (TS),
the roughly circular boundary seen about 1000~AU from the star in Figure~2.
At the TS, the stellar wind is heated, compressed, and decelerated.
The stellar wind cools adiabatically while expanding outwards, but we do
not allow it to cool below 3~K (the cosmic background radiation
temperature).  Beyond the TS there is a parabolic-shaped boundary
visible in Figure~2 separating the plasma flows of the
stellar wind and ISM, the ``astropause'' (analogous to ``heliopause''),
which extends beyond the field of view in the downwind direction (to the
left).  There is no bow shock beyond the astropause in the upwind
direction (to the right in Figure~2) like there is for the heliosphere
since the ISM flow is not supersonic.

\section{Comparing the Data with Model Predictions}

     Figure~3 shows traces of H~I density, temperature,
and velocity for the line of sight from the star toward the Sun based on
the three models listed in Table~1.  We assume that the Mg~II temperature
and velocity are identical to those of H~I, and we compute the Mg~II
density from the H~I density assuming solar Mg abundances (Anders \&
Grevesse 1989) and assuming that Mg~II is the dominant ionization state of
Mg.  We can then compute the predicted astrospheric Mg~II absorption
for all the models.
\begin{figure}[p]
\plotfiddle{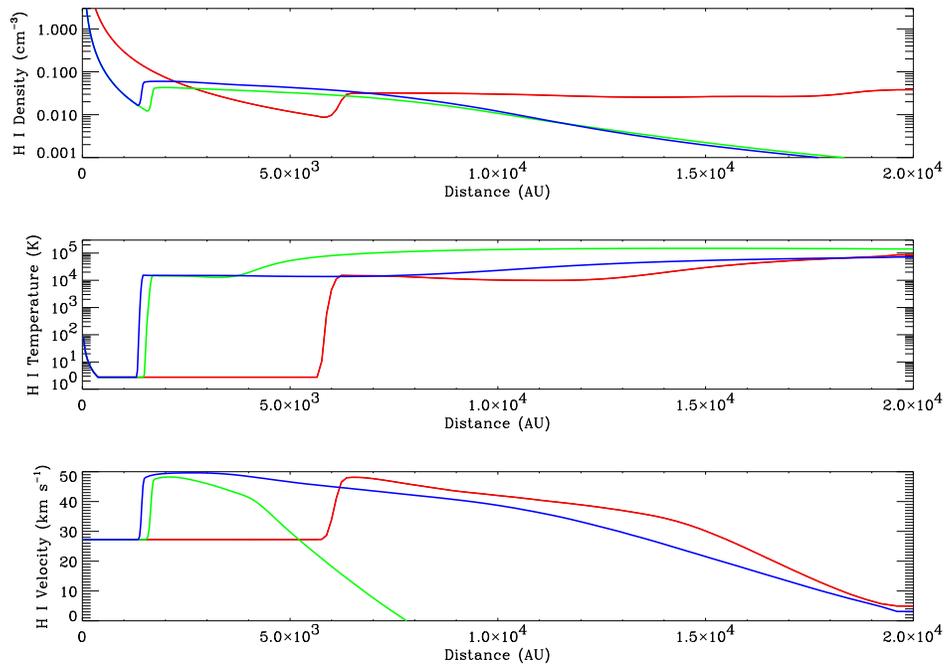}{3.1in}{90}{52}{52}{203}{-15}
\caption{Traces of H~I density, temperature, and
  velocity along the line of sight from $\alpha$~Tau toward the Sun for
  Model~1 (green lines), Model~2 (red lines), and Model~3 (blue lines).}
\end{figure}

     We find that the Mg~II column densities predicted by the models
are too low by an order of magnitude.  Figure~4 shows the predicted
absorption if we arbitrarily increase the Mg~II opacity by a factor of 10
for the models.  Despite the order of magnitude underestimate of the
Mg~II column by these initial models, they {\em do} predict the presence
of an absorption feature of about the right width at roughly the location
of the observed absorption that has been proposed to be astrospheric.
Therefore, the astrospheric interpretation of the absorption still shows
promise.  In this interpretation, the material responsible for the
absorption is the heated, compressed, and decelerated stellar wind
material outside the termination shock (TS), about $1500-4500$~AU from the
star for Model~1, $6000-15,000$~AU for Model~2, and $1500-15,000$~AU for
Model~3 (see Fig.~3).
\begin{figure}[t]
\plotfiddle{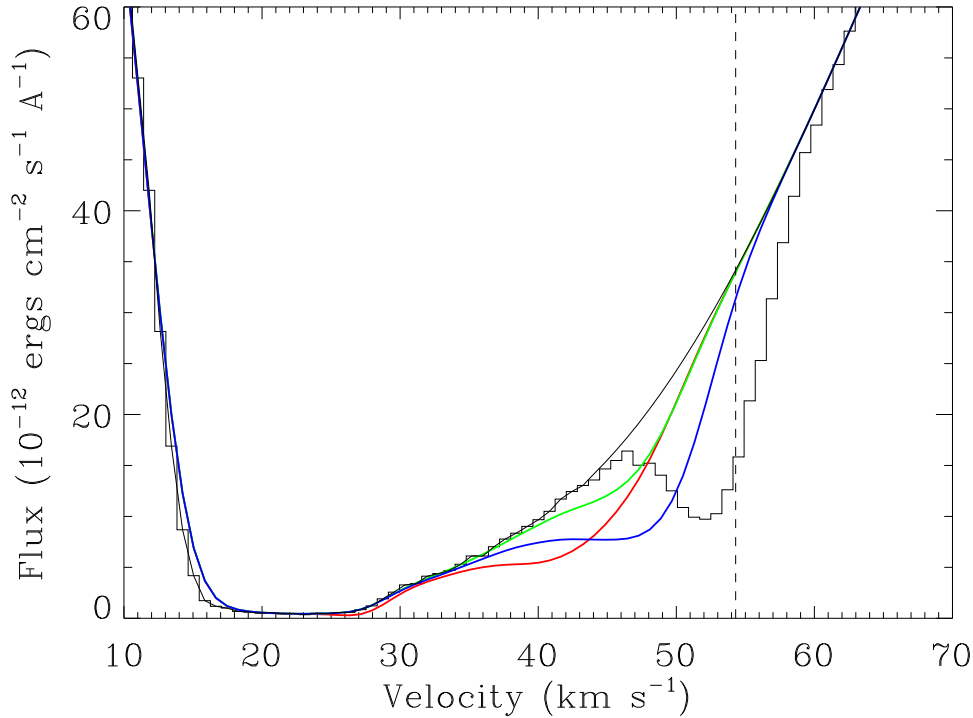}{3.6in}{90}{60}{60}{227}{-10}
\caption{Comparison of the astrospheric Mg~II
  absorption predicted by Model~1 (green line), Model~2 (red line), and
  Model~3 (blue line) with the observed absorption feature seen by
  HST/GHRS, where the Mg~II opacities of the model profiles have been
  arbitrarily increased by a factor of 10 (see text).  The vertical
  dashed line is the rest frame of the star.}
\end{figure}

     Note that if we were observing the upwind portion of the astrosphere
(i.e., $\theta<90^{\circ}$), the path through the heated post-TS material
would be much shorter and the astrospheric absorption would be much weaker
(see Fig.~2).  The downwind orientation of our line of sight through the
$\alpha$~Tau astrosphere may be the main reason we see the
astrospheric absorption for $\alpha$~Tau but not (so far) for other red
giants.  One mystery we cannot yet explain is why the astrospheric
aborption seen in Mg~II is not also seen in the O~I $\lambda$1302 and
C~II $\lambda$1335 lines observed by HST.  These lines have profiles very
similar to the Mg~II line in Figure~1, but without the narrow absorption
feature that we believe is astrospheric (Robinson et al.\ 1998).

     The model that appears to work best so far is Model~3, which predicts
a greater amount of absorption thanks to the more downwind line of sight
suggested by the assumption of the LSR for the ISM vector (see \S2).
The predicted absorption of all models in Figure~4 is blueshifted
from its observed position by about $5-10$ km~s$^{-1}$, meaning that the
models are predicting too little deceleration at the TS.  Increasing the
deceleration would lead to higher densities outside the TS, thereby also
helping to correct the problem of underpredicting the Mg~II opacity.
Unfortunately, the Model~2 experiment shows that simply increasing the
stellar wind density does not help much.  Further experimentation with
varying other model parameters listed in Table~1 is
necessary to see if a model can be found that will increase both the
deceleration at the TS and the total Mg~II column density to the required
extent to match the data.  If ultimately successful, we can then
see what constraints the observed absorption can place on the properties of
the stellar wind and surrounding ISM.

\acknowledgments

We would like to thank P.\ Frisch for helpful discussions.  Support for
this work was provided by NASA grant NAG5-9041 to the University of
Colorado.

\end{document}